\documentclass{article}

\usepackage{PRIMEarxiv}

\usepackage[utf8]{inputenc} 
\usepackage[T1]{fontenc}    
\usepackage{hyperref}       
\usepackage{url}            
\usepackage{booktabs}       
\usepackage{amsfonts}       
\usepackage{nicefrac}       
\usepackage{microtype}      
\usepackage{lipsum}
\usepackage{fancyhdr}       
\usepackage{graphicx} 
\usepackage{subcaption}
\graphicspath{{media/}}     

\title{ Trends in urban flows: A transfer entropy approach
}

\author{
  Roberto Murcio \\
  School of Social Sciences, Birkbeck, University of London \\
  Centre for Advanced Spatial Analysis, University College London \\
  \texttt{r.murciovillanueva@bbk.ac.uk} \\
   \And
  Balamurugan Soundararaj \\
  City Futures Research Centre, UNSW Sydney \\
  \texttt{s.bala@unsw.edu.au} \\
}

\begin{document}
\maketitle

\begin{abstract}
The accurate estimation of human activity in cities is one of the first steps towards understanding the structure of the urban environment. Human activities are highly granular and dynamic in spatial and temporal dimensions. Estimating confidence is crucial for decision-making in numerous applications such as urban management, retail, transport planning and emergency management. Detecting general trends in the flow of people between spatial locations is neither obvious nor easy due to the high cost of capturing these movements without compromising the privacy of those involved. This research intends to address this problem by examining the movement of people in a SmartStreetSensors network at a fine spatial and temporal resolution using a Transfer Entropy approach.
\end{abstract}

\keywords{Human mobility\and Information Theory}

\section{Introduction}

Accurately estimating the volume and flow of pedestrians in cities is one of the first steps towards understanding the structure of the urban environment 
\cite{behnam1977method,kim2013estimating, Kitazawa2004, Kerridge2001}.
Human activities such as footfall are highly granular and dynamic in space and time 
\cite{Middleton2011, Middleton2010}, and estimating them with confidence is crucial for decision-making in numerous applications such as urban management, retail, transport planning, and emergency management. Detecting general trends in the inflow of people between spatial locations is neither obvious nor trivial due to the high cost of capturing these movements without compromising the privacy of those involved 
\cite{Louail2014, Steenbruggen2013, Abedi2013, Li2019}. For the last ten years, new data sources on human behaviour that were not accessible before (WiFi, GPS, Bluetooth, for example) allowed for explanation and further understanding of human drives and decisions 
\cite{Cottineau2018, Schneider2009}. They also support individual users in decision-making as more shared knowledge becomes available at people’s fingertips daily. Insights into data from mobile crowdsourced sensors demonstrate the power of interconnectivity between people, space, and time 
\cite{Yang2018,Bonson2019,Lera2017,Graells-Garrido2018,Daamen2004}. The characteristic finding in all these studies is that people do not move randomly, i.e., following a Levy flight 
\cite{Brockmann2006}. Instead, there is strong evidence for a conserved quantity in human mobility. \cite{Louail2014, Gonzalez2008, Liu2018, Noulas2012, Song2010, Alessandretti2016}. 
All these models depend on the high granularity of their base data, allowing them to track devices through small areas (cellular towers) or at high detail (GPS).
In this work, we also studied the movement of people, but without having the explicit route followed by each device. Using aggregated counts derived from passive Wi-Fi signal probing captured across Great Britain \cite{Longley2018}, we analysed the reciprocal influence—or information transfer—between different pairs of locations. We proposed a complexity path measure that depends not only on urban morphology but also on the type of activity in each area and, consequently, on the time of day this measure is taken. A key aspect of this study is that the temporal resolution of the data employed is sufficiently granular to detect general patterns throughout the day without compromising people’s privacy.

Our three main findings are that these data types are not random and capture local activity during the day. We proposed an urban classification based on FF and a complexity index to characterise local routes.

\section{Data Description}
\begin{figure}
\begin{subfigure}{0.5\textwidth}
\includegraphics[width=0.9\linewidth]{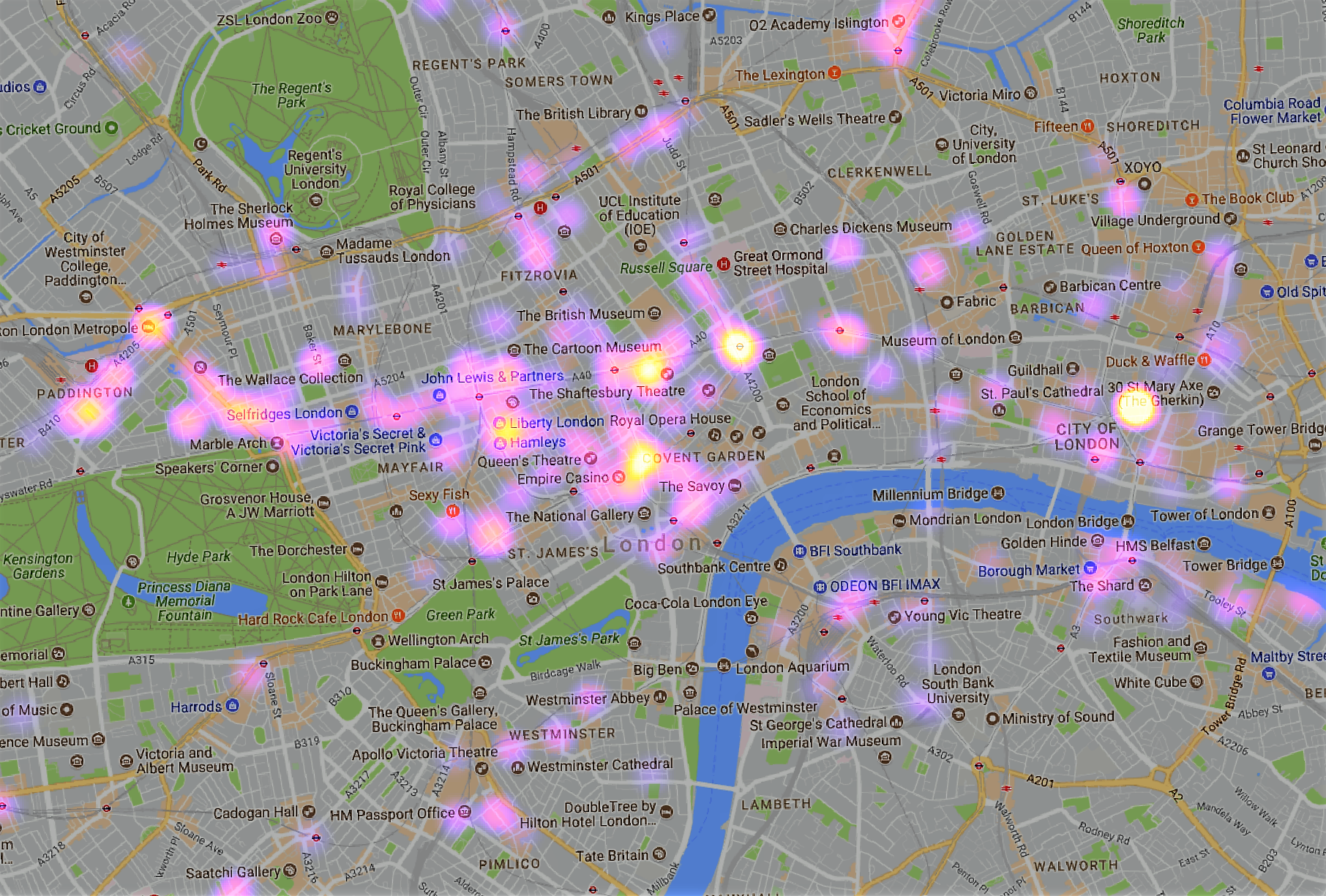}
\caption{Central London}\label{fig:london1e}
\end{subfigure}

\begin{subfigure}{0.5\textwidth}
\includegraphics[width=0.9\linewidth]{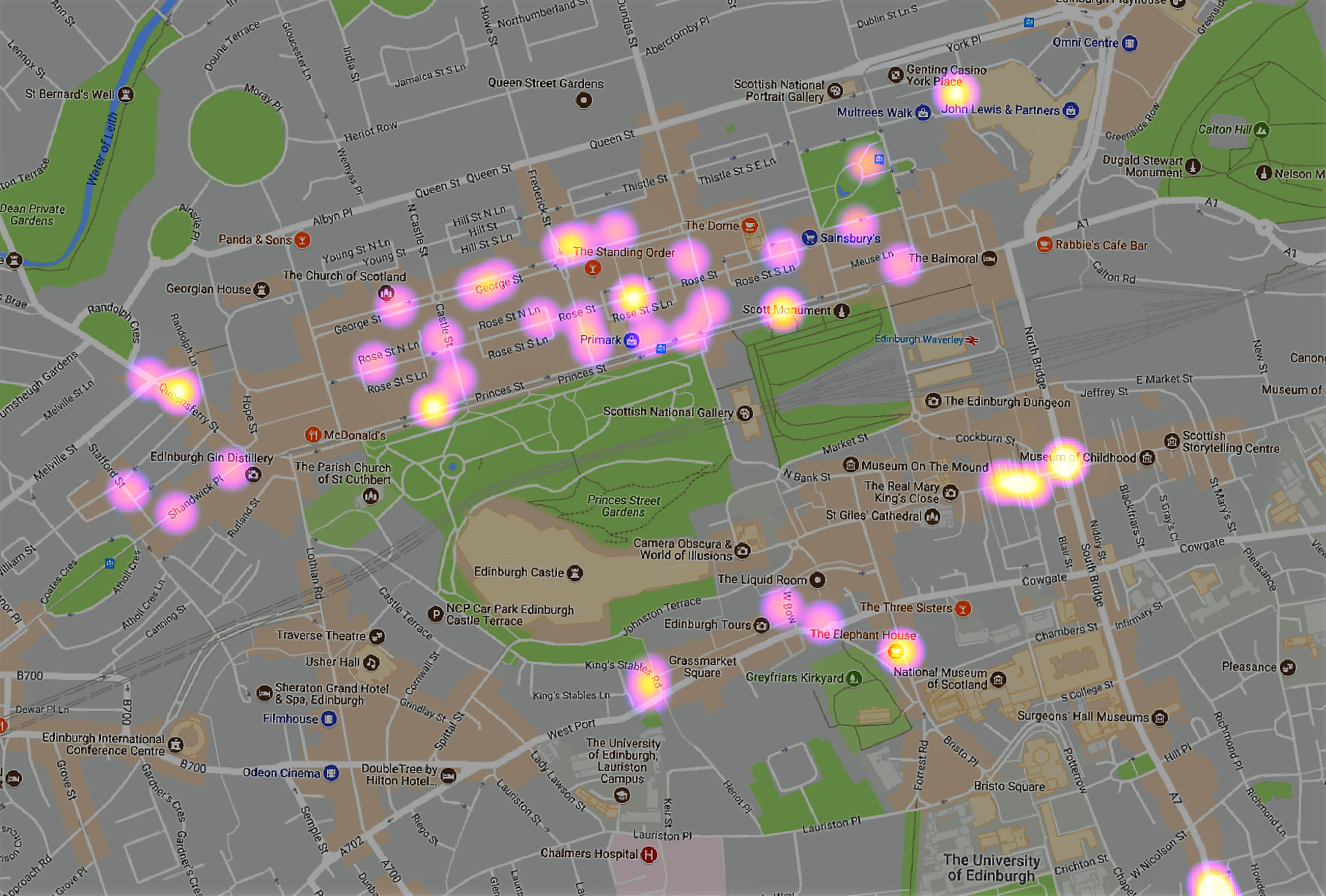}
\caption{Central Edinburgh}\label{fig:glasgowe}
\end{subfigure}
    \caption{ Distribution of sensors in a) Central London and b) Central Edinburgh. The brightest areas correspond to cities with a higher number of  sensors in operation  (which does not necessarily imply a higher FF, although it is the case in London.}
    \label{fig:locations_heatmap_uk}
\end{figure}

The data for this research form part of the Smart Street Sensor project \cite{cdrc} - a large-scale study commissioned by Consumer Data Research Center, UCL, to understand footfall (FF) in retail centres across the UK.
The project employs a proprietary sensor network that records mobile device signals to discover available Wi-Fi access points. Once captured, the data is processed into five minutes aggregated counts.  The precise algorithm followed to map Wi-Fi probe requests into FF counts and the associated bias are described in section \ref{SupM:S1} in the Supplementary Materials. For the work presented here, we used data from the 1st of January 2017 to the 31st of December of the same year across 95 cities in Great Britain. Figure \ref{fig:locations_heatmap_uk} shows an example of the sensor's locations in Central London and Central Edinburgh, UK.

The volume of FF reported at these locations varies widely. In 2017, the busiest location was inside Waterloo train station, London, with an average of 500 people per hour, while the quietest one was at Rockingham Road, Market Harborough, with only five people on average passing by this location.  

\subsection{Footfall signals}
\begin{figure}
\centering\includegraphics[width=1\textwidth]{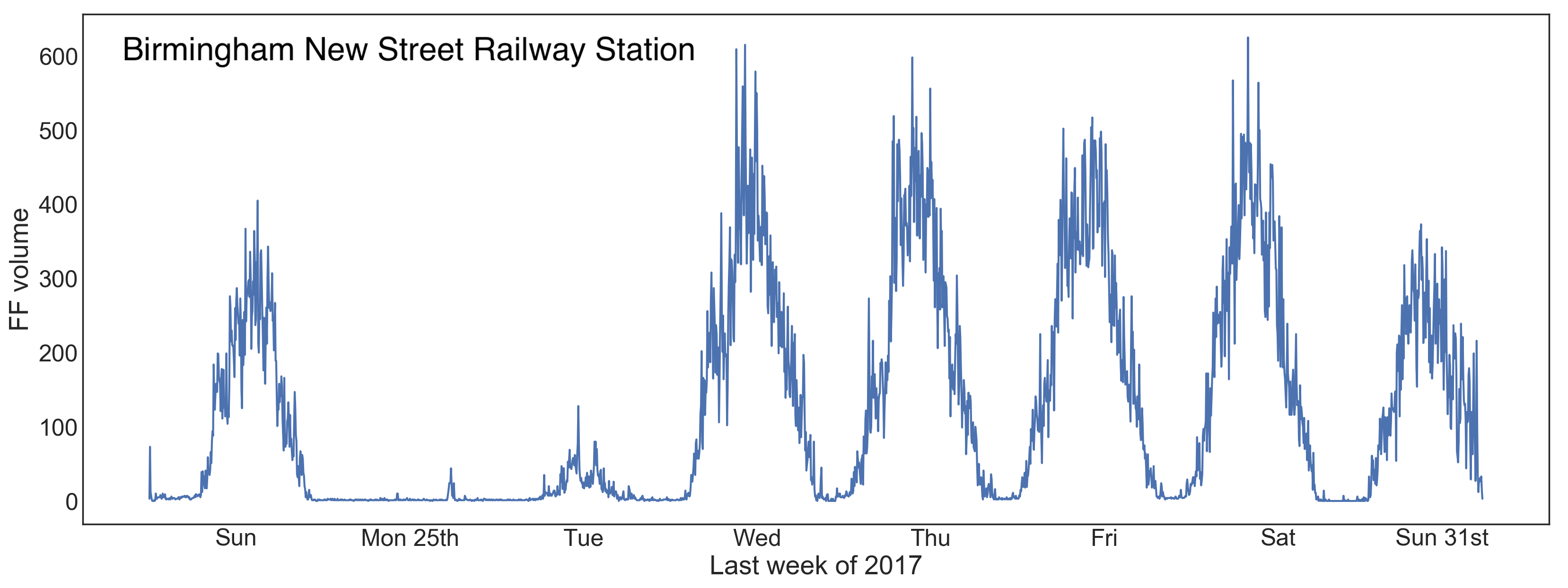}
\caption{FF signals exhibit weekly circadian rhythms. For this particular location -inside Birmingham New Street rail station- we can observe an early peak in FF volume and high volume every five minutes during the middle of the day. This signal also reflects the lack of activity around the station on Christmas and boxing}
\label{fig:ffsignal}
\end{figure}

Each location generates a daily time series – FF counts as a function time – or FF signal,  that captures the local movements of devices around each sensor (Fig. \ref{fig:ffsignal}). The FF signal is the basic unit of information in this study. 

\begin{figure}
\centering\includegraphics[width=1\textwidth]{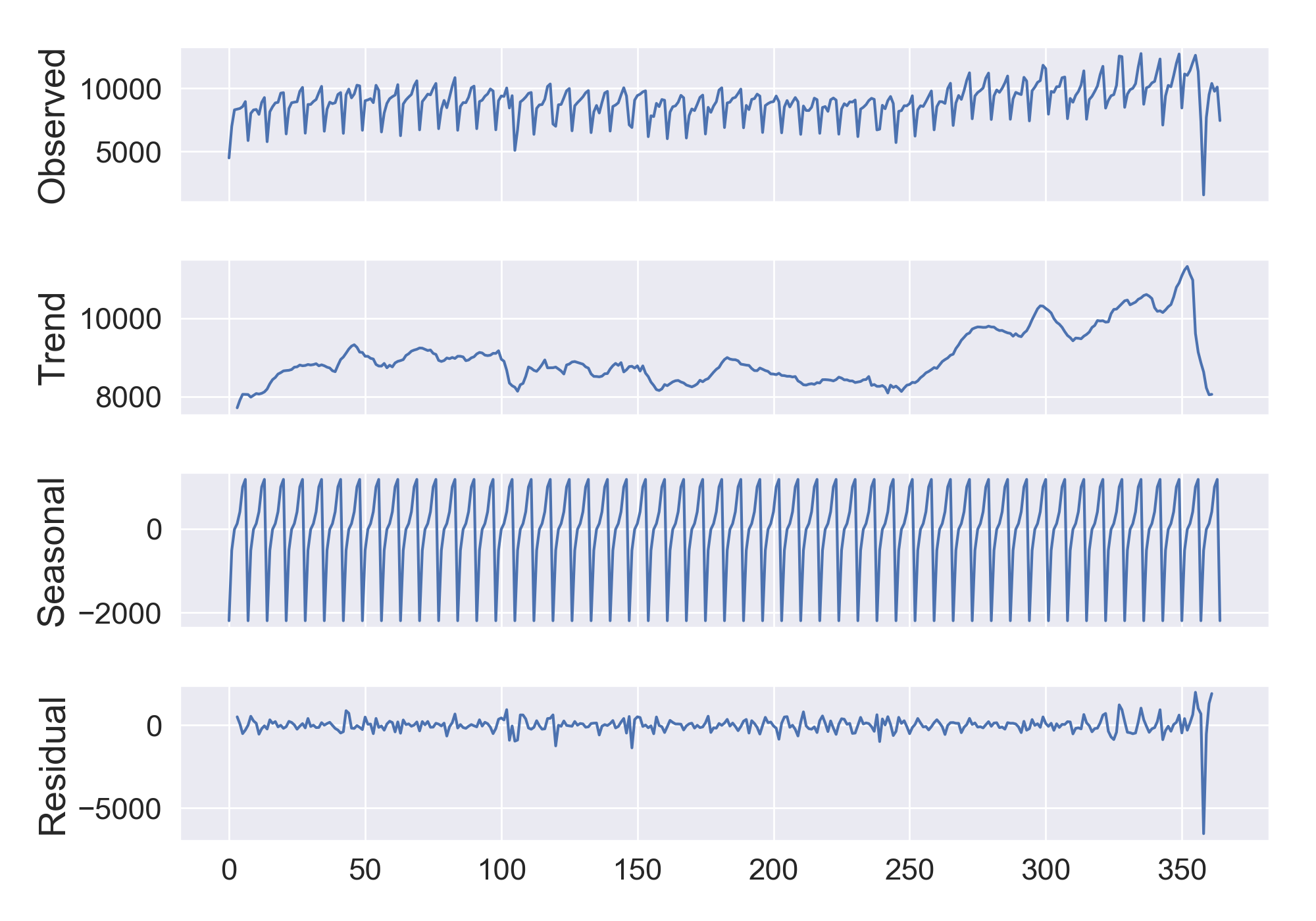}
\caption{National average footfall in 2017 - decomposed time series}
\label{fig:decomp_ts}
\end{figure}

To introduce the general characteristics of the FF signals, an average FF was computed for all the five-minute periods in 2017 across all the locations in Great Britain, and the resulting aggregated time series was decomposed into standard trend, seasonal and residual components (Figure \ref{fig:decomp_ts}). Since the magnitude of the seasonal variation appears to be constant, the additive model was used to describe the time series.

The FF signal's trend component shows volume growth throughout autumn, leading to peak activity during the festive pre-Christmas period. A sharp drop in FF near the end of the year is due to substantially lower counts on Christmas Day compared to the previous days and the next days. FF also exhibits a clear weekly seasonal pattern. Workdays are typically busier than weekends, and Sundays tend to record weekly lows. The residual component is expectantly present, as the probe requests used to construct the FF counts are rather noisy, initially coming with many sources of measurement error that had to be accounted for through data cleaning (Supplementary Information \ref{SupM:S1}). However, the change in the residual component does not exhibit any particular pattern, except for displaying the sharp drop near the year-end, which was already attributed to the low footfall on the 25th of December.

\subsection{Assessment}
The FF signal's trend and seasonal components indicate this data has a non-random nature and that it detects the ups and downs in FF activity around different urban locations. To further assess these features, we first look if these data are capturing typical human activity patterns, and second, we use as a proxy three different sources: 1) extraordinary events, 2) public transport figures, and Google popular times \cite{gapi} and 3) compare the FF signals with previous results obtained from similar sources, like mobile phone data.

\begin{figure}[ht]
\centering\includegraphics[width=1\textwidth]{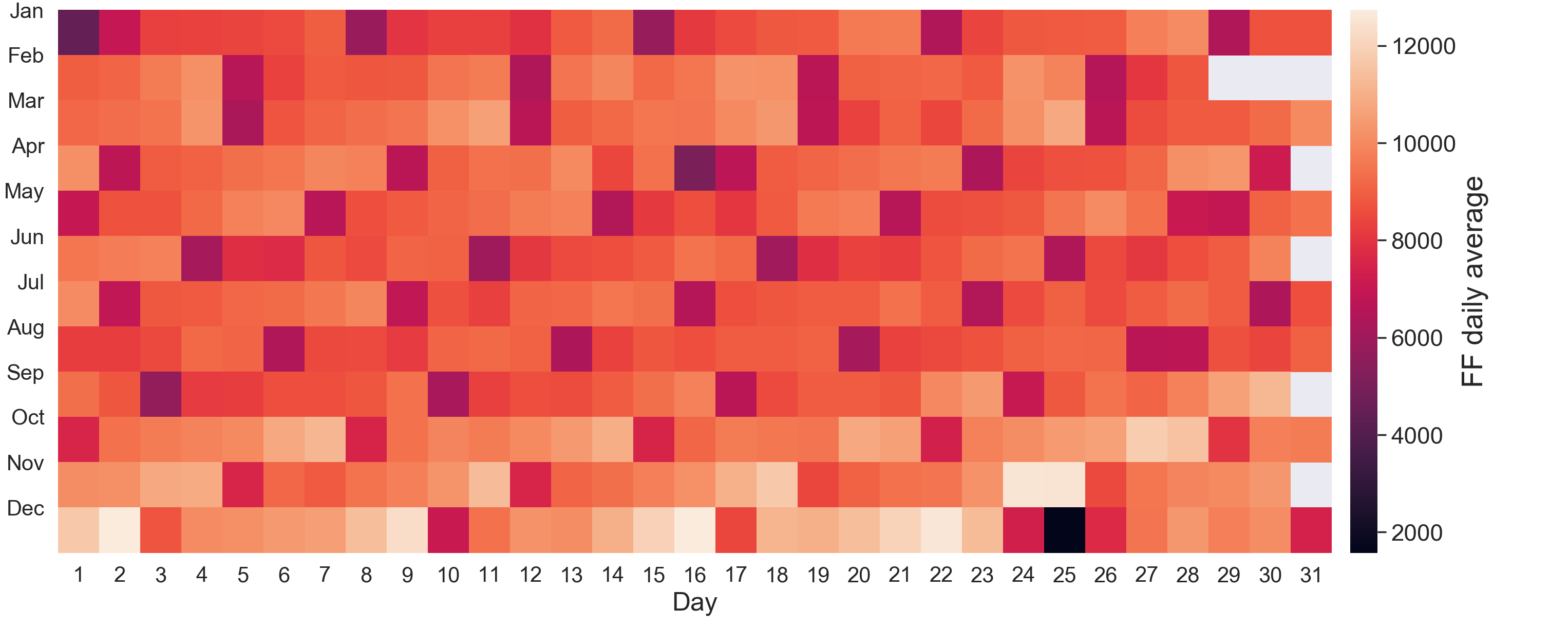}
\caption{Monthly FF counts. There is an evident drop in activity on the 25th of December and the 1st of January, while the peak in FF was reached on the 16th of December (a Saturday two weeks from Christmas) }
\label{fig:yearly_footfall}
\end{figure}

Figure \ref{fig:yearly_footfall} shows the daily average change in FF for all the locations in Great Britain. We can see how FF remains quite stable throughout the whole year, in fact, throughout each month and each week with some clear exemptions: The 1st of January and the 25th of December are the lowest points in FF (black squares), while  December the 2nd, 9th, 16th, 22nd and 23rd (Saturdays and Fridays before Christmas's day) are quite busy as expected. The weekly rhythm is also evident. The blocks between Monday and Sunday are easily identified in each Month; for example, in October, we can trace a week between red blocks (October 1st was a Sunday) and see how each day of the week shares similar palette tones, i.e., shares similar FF.

\begin{figure}
\centering\includegraphics[width=1\textwidth]{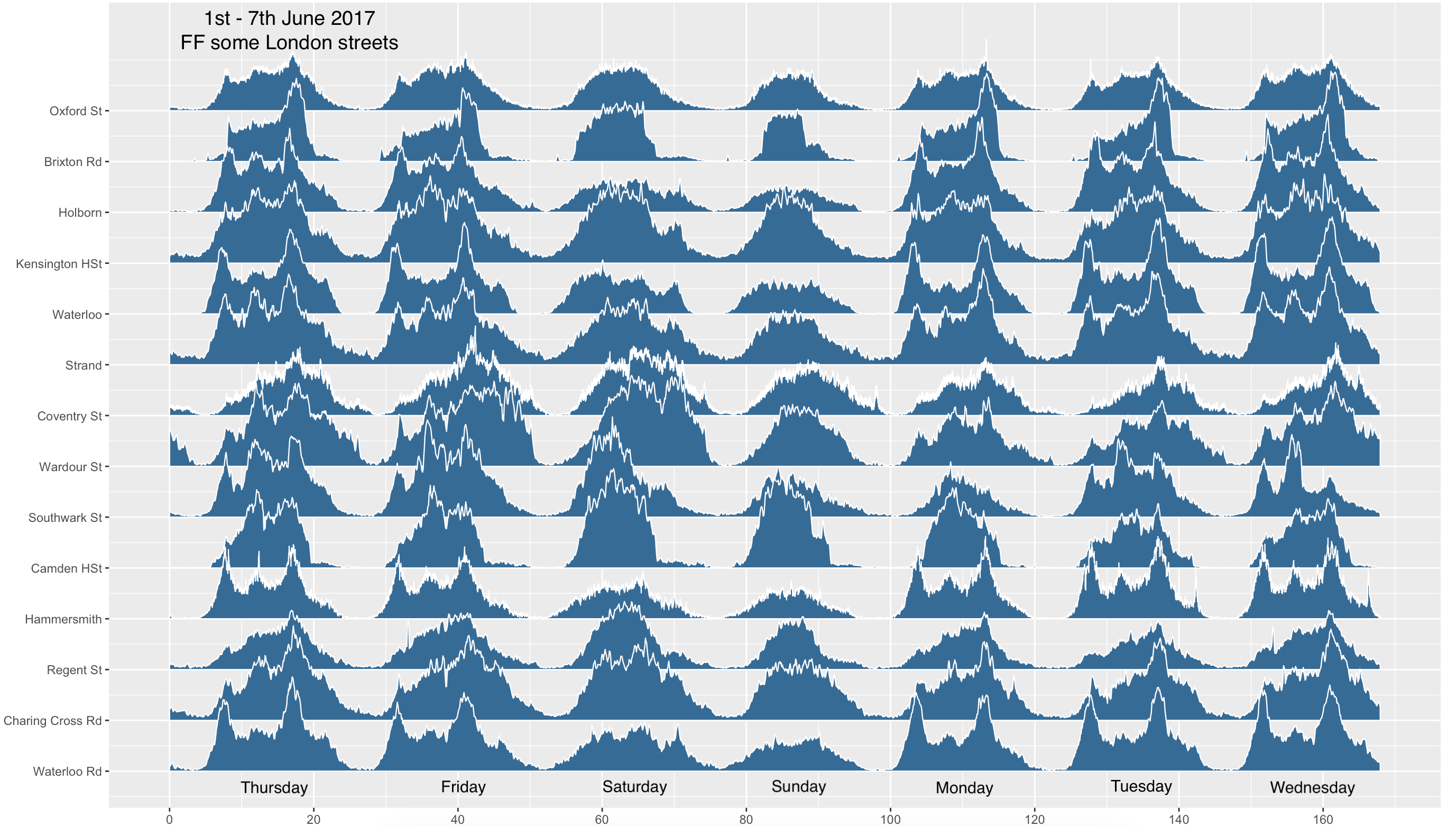}
\caption{FF signal at some London streets during the day}
\label{fig:signalsLondon}
\end{figure}

The next step was to investigate the daily variation in FF across different locations.  Figure \ref{fig:signalsLondon} shows a week of FF signals at fourteen London locations. As in Figure \ref{fig:yearly_footfall}, the differences between weekends and weekdays are present.  The shape on weekends tends to be more like a plateau, while weekdays exhibit one, two, or three clear peaks. Take, for instance, Brixton Rd, a typical London weeknight location, as it has a peek at its right part during weekdays, but it is a plateau during weekends. This initial visual inspection of the FF signatures demonstrated that Monday through Thursday display patterns different from those observed during the weekends, and even though Fridays are mostly similar to the rest of the workdays, some locations may have unusual activity in the evening hours. In general, the majority of locations from this data typically record busier workdays compared to weekends \cite{Longley2018}

\subsection{Detecting extraordinary events} 
\begin{figure}
\begin{subfigure}[b]{.45\linewidth}
\includegraphics[width=\linewidth]{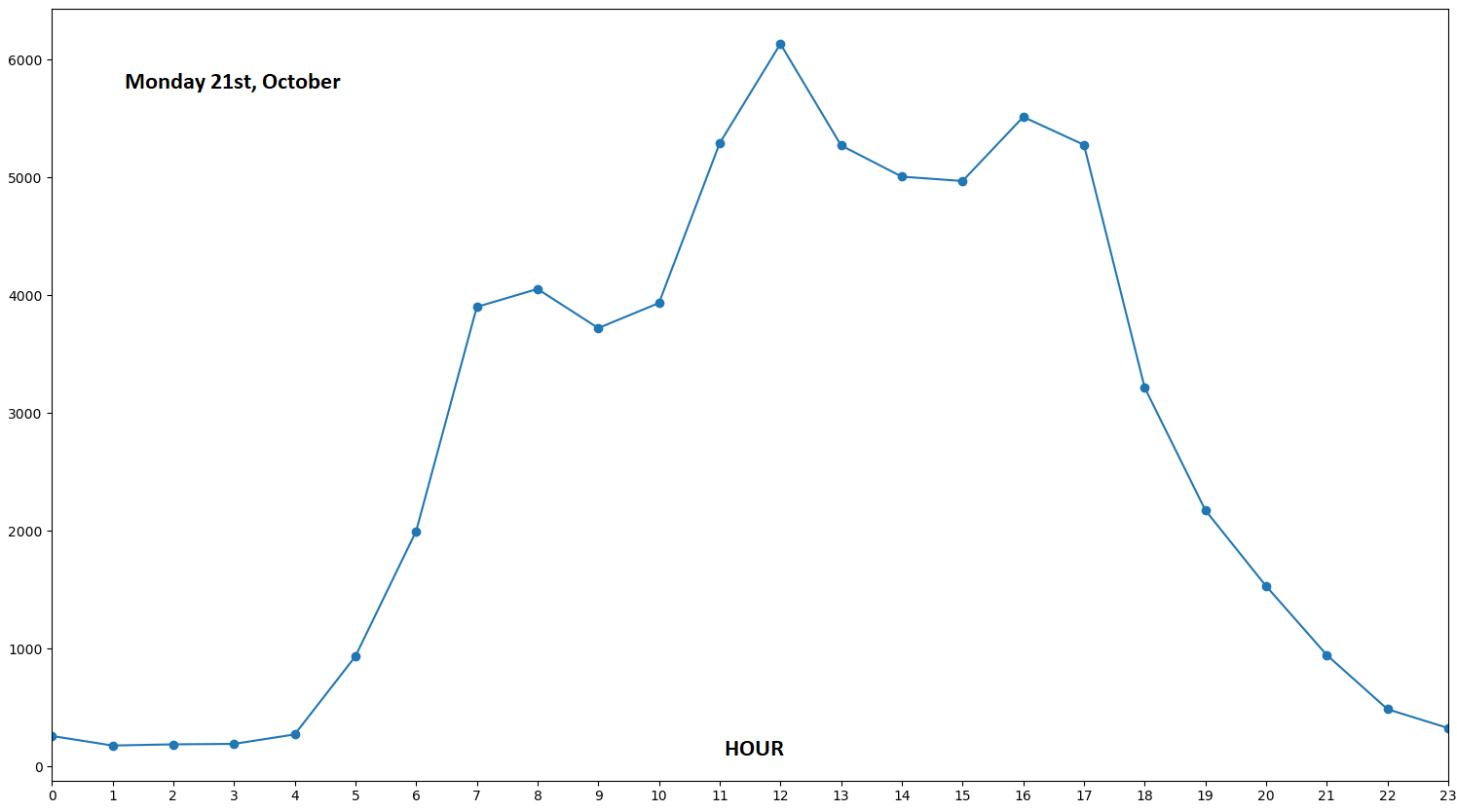}
\caption{Average Monday}\label{fig:notH1}
\end{subfigure}
\begin{subfigure}[b]{.45\linewidth}
\includegraphics[width=\linewidth]{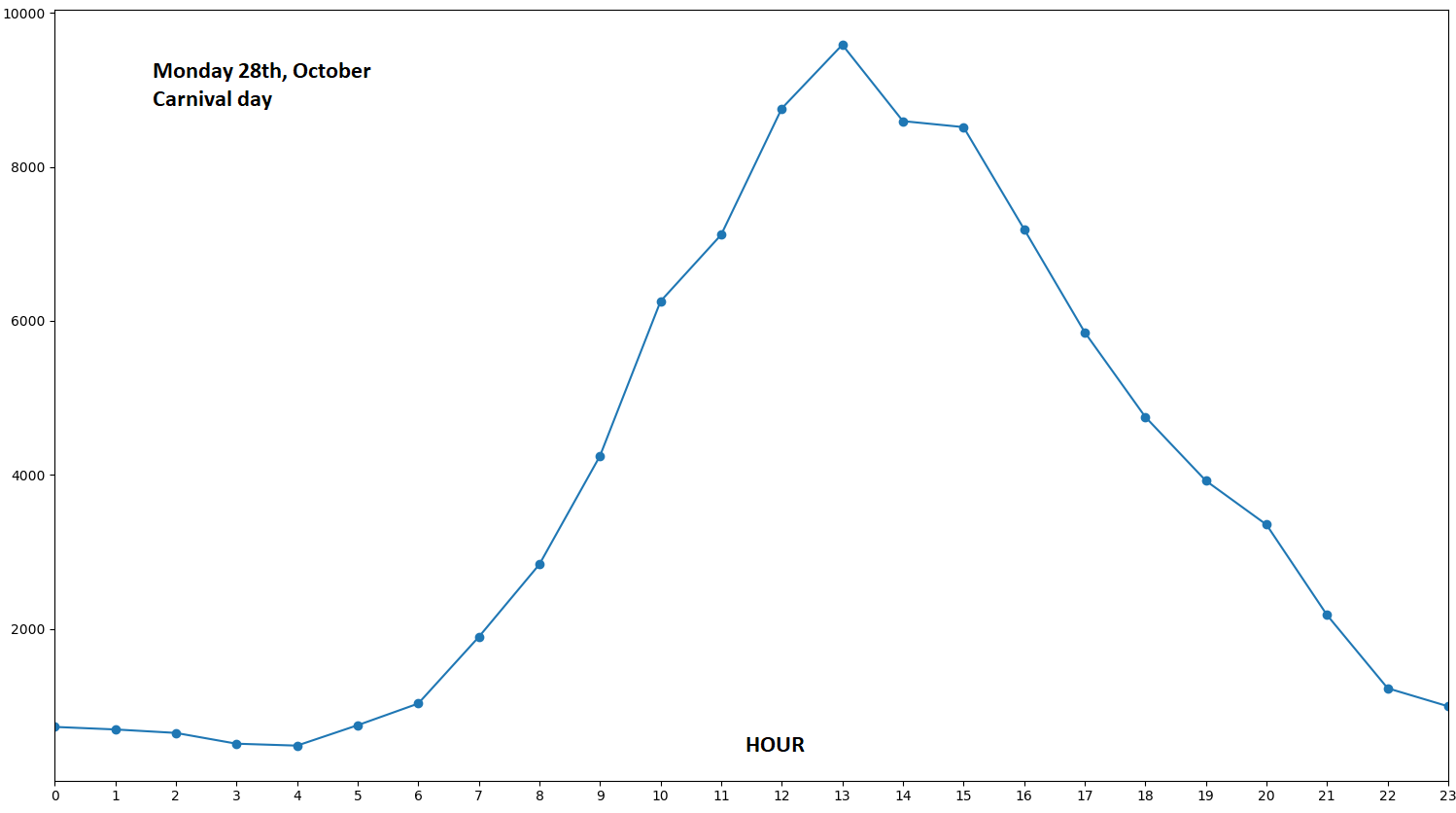}
\caption{Carnival Monday}\label{fig:notH2}
\end{subfigure}
\caption{Different signals for different activities}
\label{fig:notHill}
\end{figure}

The Notting Hill Carnival is a massive yearly event in London that attracts thousands of visitors over three days. In 2017, the carnival occurred from Saturday, 27th to Monday, 28th October. There are no sensors installed around the Nothing Hill tube station, but there are four around the High Street Kensington station, which is a popular location for getting into the event. Fig. \ref{fig:notHill} shows the change between an average Monday and a Carnival Monday. In the former,  the typical morning/afternoon peaks at rush hour, plus a peek at lunchtime, are clear, while on Carnival day, there is only one peak at 1 pm and from 9 am to 7 pm, the volume of FF is in orders of more than 4,000 people per hour. The data captured by these sensors successfully detect the change in FF around this area due to the unusual amount of activity and pedestrian flows.

\subsubsection{London underground comparison}
\begin{figure}
\centering
\begin{subfigure}[b]{.90\linewidth}
\includegraphics[width=\linewidth]{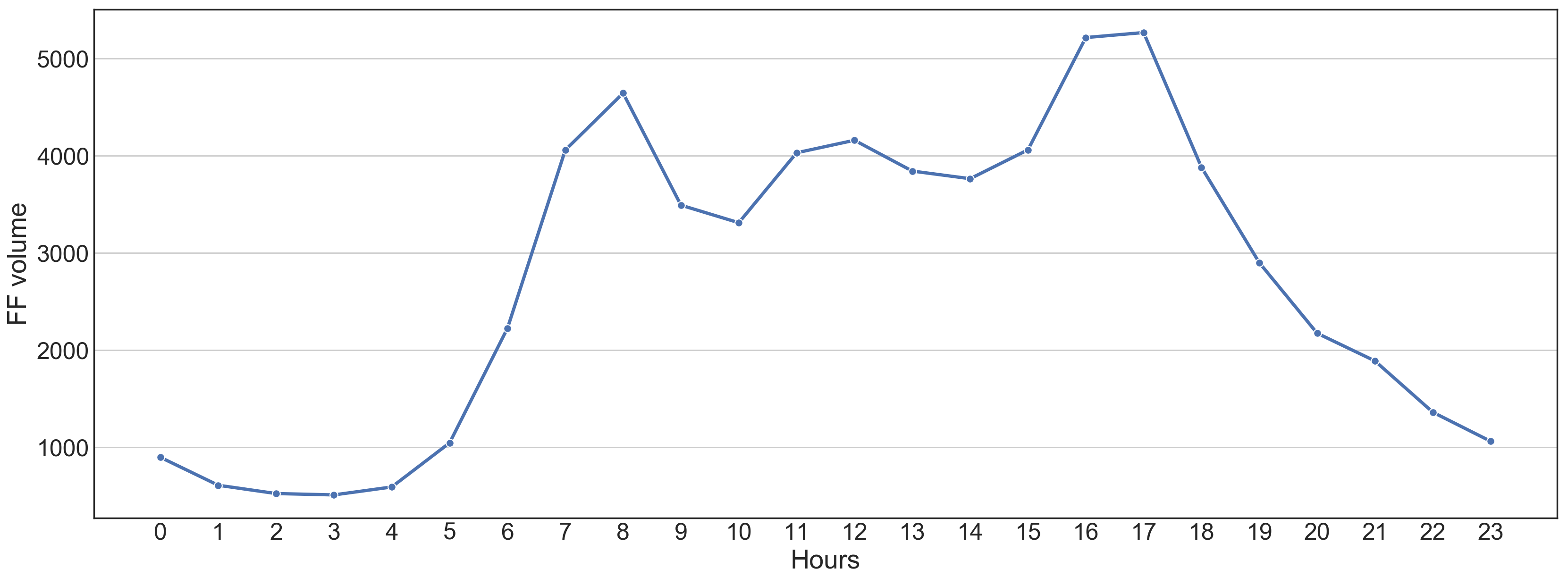}
\caption{Average weekday FF signal}\label{fig:tott}
\end{subfigure}
\begin{subfigure}[b]{.45\linewidth}
\includegraphics[width=\linewidth]{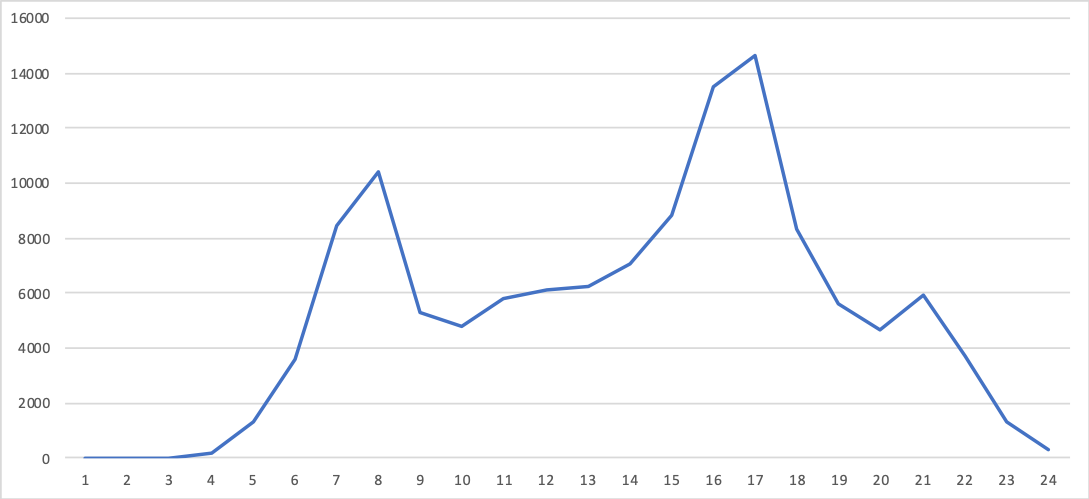}
\caption{Aggregated weekdays counts (entries and exits) at Tottenham Court Road underground station}\label{fig:totTube}
\end{subfigure}
\begin{subfigure}[b]{.45\linewidth}
\includegraphics[width=\linewidth]{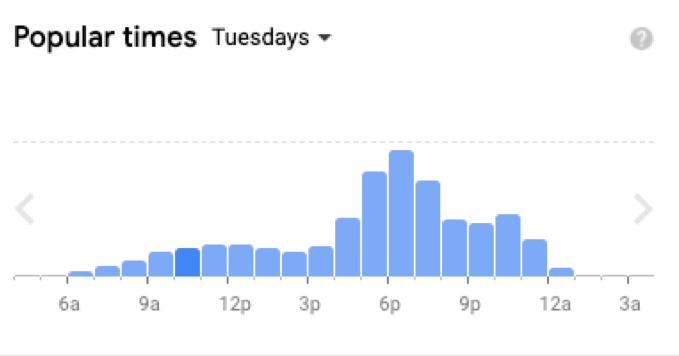}
\caption{Screen capture of Google popular time plot on a weekday in November}\label{fig:popTimes}
\end{subfigure}
\caption{Comparative plots for three sources of peoples' counting measures around Tottenham Court Road underground station}
\label{fig:totcompare}
\end{figure}

If these sensors are capturing the activity trends around different areas, they should be in some accordance with other counting people sources, like the Entry and Exit figures for each underground station reported by the Transport for London (TFL)  authority or Google's Popular Times feature. The former (obtained from  \cite{tfl}) provides the average 15-minute aggregated counts for November 2017, while the latter can obtained by selecting particular locations in Google Maps \cite{gapi}. We selected all of London's locations at a 5-minute walking distance or less from an underground station (80 locations). Then, we selected points of interest around the stations with information on Google's popular times. We found a high Spearman correlation among FF signal-aggregated counts-Google times for practically all locations (see SI.b for more information). Figure (Fig. \ref{fig:totcompare} ) shows an example of this validation for a location around the Tottenham Court Road underground station.

\subsubsection{Mobile phone data comparison}
\begin{figure}
\centering\includegraphics[width=\linewidth]{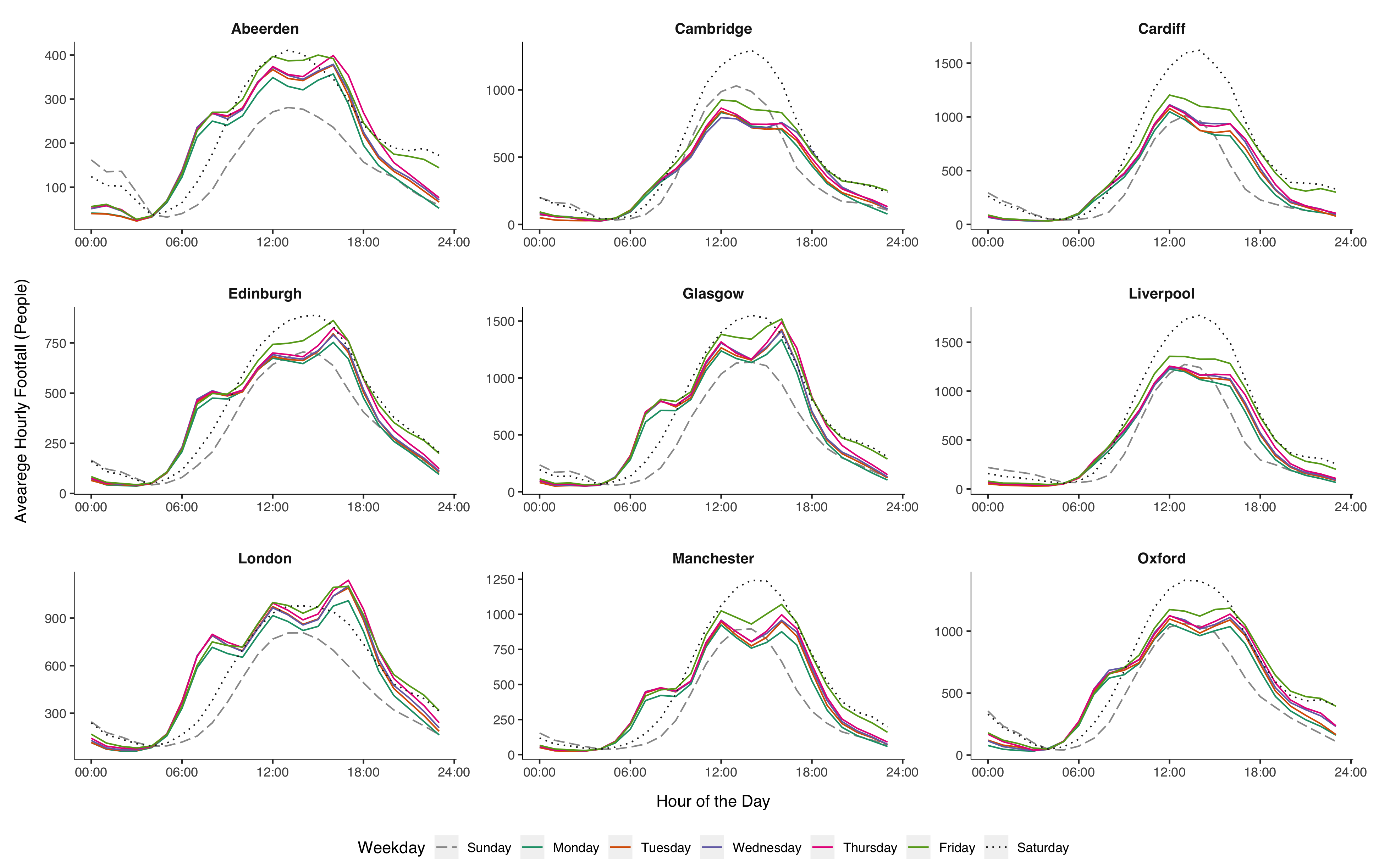}
\caption{Daily FF signals for six UK cites}
\label{fig:ffcities}
\end{figure}

Finally, we compared the FF signals with one of the many previous results obtained from mobile data. Figure 4 from \cite{Louail2014} shows how six Spanish cities behave similarly in terms of the distribution of the number of users at different times of the day and how the morning and afternoon peaks are marked. In Figure \ref{fig:ffcities}, we reworked the same plot but for six UK cites, obtaining a similar daily behaviour, i.e., the same morning and afternoon peaks. One key difference is that the FF data also detects the lunch peak. 

The above comparisons provide enough evidence that these data are not random and can be used as a proxy to detect human activities.

\subsection{Transfer Entropy TE}
\subsubsection{Information measures}
Shannon's entropy represents the basic measure of information and is the preferred measure for detecting the reduction in uncertainty by any measurement $x$ of a random variable whose probability is $p(x)$.\\

\begin{equation}
H(X) = - \sum_{x\in X}p(x)log(x), \sum_x p(x)=1
\label{eq:shannon}
\end{equation}

Extending Shannon entropy to measure the uncertainty between two interacting random variables X and Y gives rise to the concept of Mutual Information
\begin{equation}
I(X, Y)=\sum_{x\in X}\sum_{y\in Y} 
p\left(x,y\right) \log \left(\frac{p\left(x,y\right) }{p(x) \cdot p(y)}\right)
\label{eq:MI}
\end{equation}

In the complex urban landscape, nearby locations have some degree of influence on each other. It is then plausible that a pedestrian X, counted at location $S1$ at time $t$ is the same pedestrian X counted at location $S2$ at time time $t+1$. However, there is no route information in the FF data (as in \cite{Louail2014, Gonzalez2008, Lee2017}) and with this data, there is no reliable procedure to detect such situations. Instead, we propose that measuring the causal influence that a location may have over another during a day will provide some insights into the pedestrian directional flows at a local scale.

This casual influence can be studied as the exchange of information that the FF signal at \textit{A} has over \textit{B} and vice versa.  Here, we measured this effect using the Transfer Entropy (TE) \cite{Schreiber2000}. 
TE is an information measure that informs about the future state of a random variable $Y$, taking into account not only the past of $Y$, but the past of another random variable $X$ as is a well-established technique used to measure the transferring of information between different layers or actors in many complex systems, 
such as  neural connectivity \cite{Timme2018,vicente2011transfer}, 
 urban studies \cite{Murcioa}, 
 financial markets \cite{Dimpfl2013,marschinski2002analysing}, social media \cite{Borge-Holthoefer2016} and thermodynamics \cite{Prokopenko2013}. 
 
In TE, we have two sample spaces of information, $X = {x1, x2,…, xt}$ and $Y = {y1, y2,…, yt}$, the transfer entropy from $X$ to $Y$ is obtained from defining the entropy rate between two systems as the amount of additional information gained from the next observation of one of the two systems as :

\begin{equation}
T(Y, X)=\sum_{t=1} p\left(x_{t+1}, x_{t}, y_{t}\right) \log \left(\frac{p\left(x_{t+1}, x_{t}, y_{t}\right) \cdot p\left(x_{t}\right)}{p\left(x_{t}, y_{t}\right) \cdot p\left(x_{t+1}, x_{t}\right)}\right)
\label{eq:TE}
\end{equation}

Where $x \in X and y\in Y$  \textit{t} indicates a particular temporal resolution, Eq. \ref{eq:TE} measures the reduction in uncertainty at $y_t$, given $x_t$ and $y_{t-1}$ in comparison with the case when only $y_{t-1}$ is known. Clerly, $T(Y,X)\neq T(X,Y)$. In the context of this paper, $X$ is the FF signal obtained at a location \textit{A}; $Y$ is the FF signal obtained at a location \textit{B}; and \textit{t} is a 5-minute bin running for an entire day.

\section{Results}

Our results are based on 53 million anonymous records corresponding to all the aggregated 5-minute Wi-Fi counts in 2017.

\subsection{Correlation}
To make sense of the complex interactions between locations, we first look at the daily correlation between all pair of locations $S1,S2$ that are at a 5 min walking distance from each other. This distance is measured using the latitude and longitude of each location as an input for the Google walking directions API \cite{gdmatrix}. For example, Fig shows the FF signal of pair of locations with 5-miniutes.

\begin{figure}[h]
\centering\includegraphics[width=0.8\textwidth]{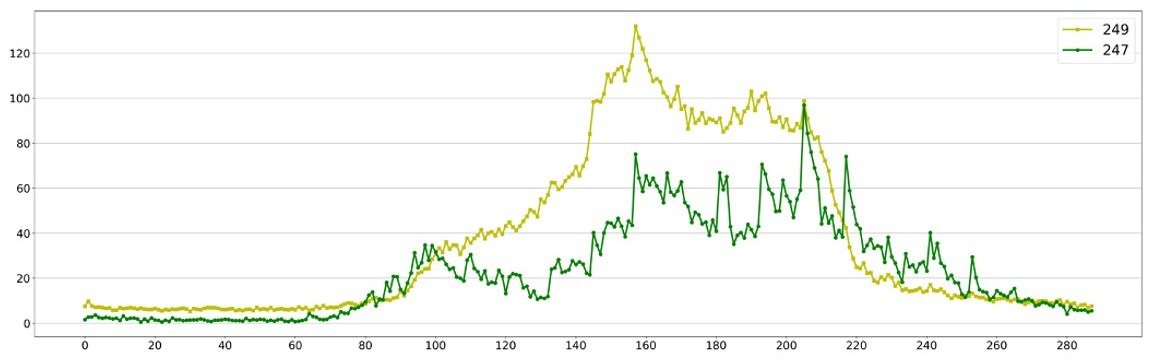}
\caption{Example of the FF signal of two Central London locations. They exhibit a high correlation in the hourly counts, suggesting that $S1$ has the a similar  mobility behaviour at the same hour than $S2$}
\label{fig:exs1s2}
\end{figure}

\begin{figure}[h]
\centering\includegraphics[width=0.8\textwidth]{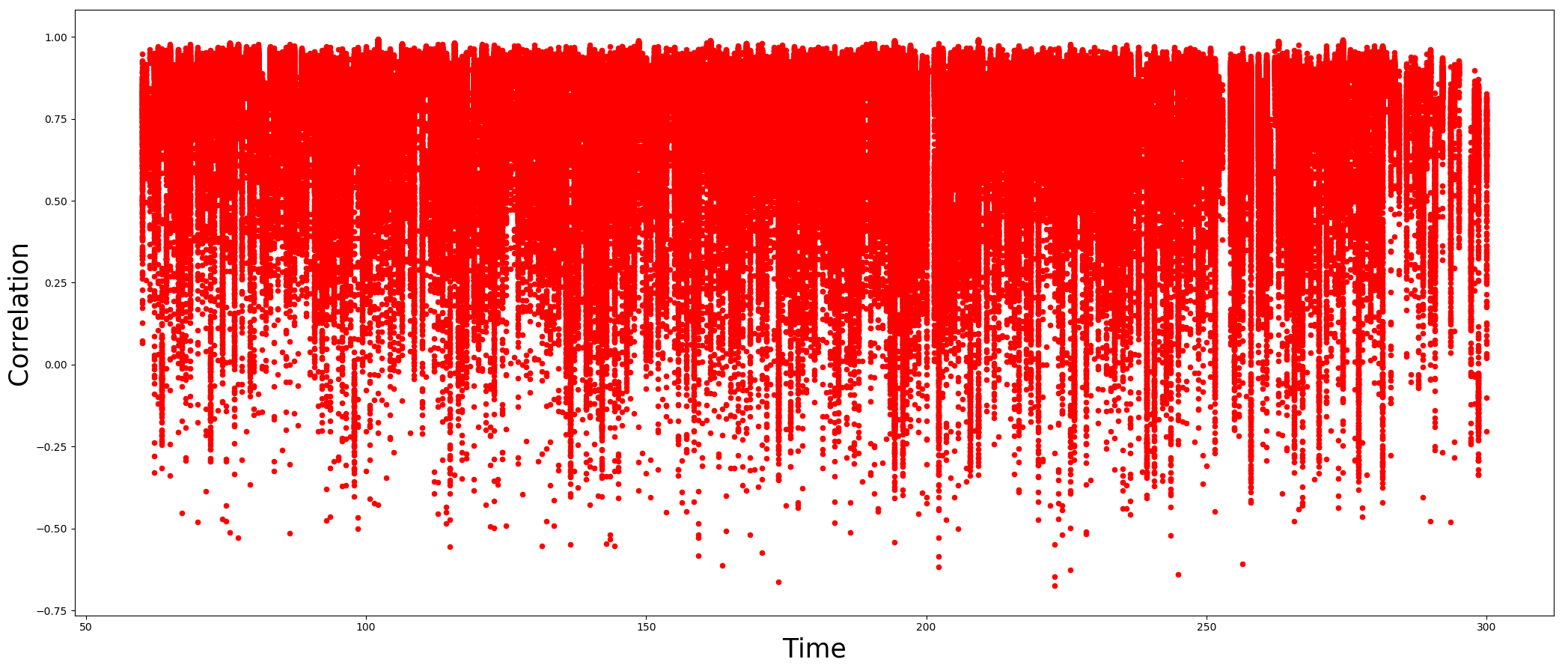}
\caption{Relation correlation-distance}
\label{fig:scatterTC}
\end{figure}

To make each pair comparable, we selected those locations with a similar number of daily measures. After this, the final number of pairs studied was 842. Figure \ref{fig:scatterTC} shows the scatter plot between the Pearson's correlation  vs the walking distance (in seconds) for all these pairs. As we can see, distance is not a factor in the value of the correlation, nor is the urban configuration between pairs of sensors, as we can find high/low correlations over the whole 5-minute range. Also, we observed a denser concentration of points on the range of correlations $> $0.5, with a decrease after the second 280 (Figure \ref{fig:scatterTC}). 

To define areas with high/low correlation and time, we divided Figure \ref{fig:scatterTC} into four quadrants with centre at Time=150 seconds and Correlation=0.5\footnote{Quadrants are labelled in the Cartesian usual way: I at the right top, II at the left top, III at the left bottom and IV at the right bottom}. Depending on the day, the same pair could belong to more than one quadrant; in fact, there are only 61 pairs that belong to one quadrant, 506 belong to two, 275 are located in three, and no pairs are present in four quadrants. The fact that these pairs belong to more than one quadrant reflects the complexity of people's movements in urban areas. One day, two locations could be highly correlated, while the next one, this correlation may fade. For example

We can calculate the percentage of days that each pair is in quadrant I and then take the maximum percentage by pair. The distribution of these maximums exhibits a Gaussian behaviour, with a mean at $\mu=61 \%$, so for most locations, half of the days studied belong to one quadrant and half in another. Incidentally, the quadrant with the maximum percentage (even for a small difference) at 95\% of the locations is quadrant II (locations with high correlation and at a short distance from each other). 
Again, we cannot characterise most pair locations with only the correlation value. To estimate the direction of flows, we followed a Transfer Entropy approach.

\subsection{Transfer Entropy}
TE may detect a casual interaction between $S1,S2$, even if these two locations are far away (one in London and the other in Glasgow, for example), making the obtained value irrelevant. To avoid these situations, we restricted our analysis to the 842 analysed in the previous section. 

A key element of Eq. \ref{eq:TE} is its asymmetric nature ($T(S2,S1)\neq T(S1,S2)$), which allows us to assign a direction $vF$ to the flow of information between $S1$ and $S2$ by simply comparing the TE value: $vF=1$, if $T(S2,S1) > T(S1,S2)$; $vF=2$, if $T(S2,S1) < T(S1,S2)$ and 0 if $T(S2,S1) \approx T(S1,S2)$. It could be the case that for a particular configuration, the TE value reported is not statistically significant (a negative value, for example). In this case, we define $vF=-1$.

\begin{figure}[h]
\centering\includegraphics[width=0.8\textwidth]{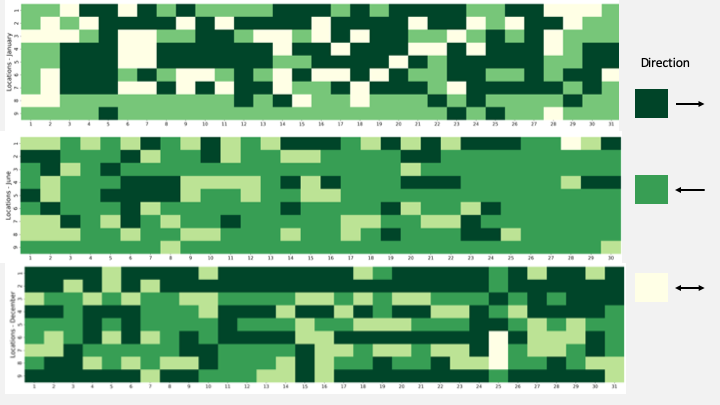}
\caption{TE between studied location pairs}
\label{fig:dFplot}
\end{figure}

Figure \ref{fig:dFplot} shows an example of $vF$  daily evolution for nine pairs during three months (2017). No pair has the same $vF$ throughout the days, and the division between Weekdays and Weekends is no longer present. We observed this anti-pattern repeat through all the pairs studied. This analysis reveals pedestrian movements' complex (almost chaotic) dynamics, as minute-by-minute micro-changes profoundly affect the overall daily $vF$ value. Nevertheless, we can recover the most typical direction for each pair during 2017.

By ranking $vF$ about its frequency, we explored if a pair $S1,S2$ has a preferred direction and if this pattern has a temporal component (pedestrian micro-dynamics in February could be quite different from the ones observed in July, for example,). If the percentage difference between the nominal value of rank one and two is greater than 10 points, then rank one is the preferred direction for that particular pair.

Finally, we look into the relationship between correlation and TE. We found that we can explain, in ~90\% of the cases, the different regimes in different weekdays for the same pair:  

\begin{enumerate}
    \item For high correlation, high TE locations, the flows between them  it is equal.
    \item For low correlation, high TE locations, the flow between them is greater from one to the other
\end{enumerate}

\subsection{Route complexity}

Both the correlation value and the TE are essentially non-spatial measures. Although we can identify locations by their correlation-distance-flow direction (add), we still don't explain why two locations with similar distance/correlation values have different TE or vice versa. To add a spatial component to our analysis, we derived a \textbf{\textit{score route}} looking into the subset of all the possible  walking routes from $S1$ to $S2$, getting these as follows:

\begin{enumerate}
   \item Get all the possible routes (and step-by-step directions) from the Google API for a total of ~2,000 routes
   \item Select the route that takes less time to have only one route per location. 
   \item Performed a semantic exploration of the walking directions to isolate different relevant words
   \item Assign a numerical value to each word, weighted by its associated distance
   \item Sum all values to compound a single score.
\end{enumerate}


We selected a set of thirteen meaningful directional words and assigned a value to each of them that tries to capture different degrees of complexity when navigating a walking route (see table \ref{table:weightroute} )

\begin{table}[ht]
\centering
\begin{tabular}{|c|c|c|} 
 \hline
Direction & Value & Context\\ [0.5ex] 
 \hline
head,toward(s),continue,follow,straight,walk & 1 & "Head $<$direction$>$ toward $<$place$>$\\
slight,sharp & 2 & "Slight right onto $<$place$>$\\
turn & 3 & "Turn $<$left,right$>$ onto $<$place$>$" \\ 
cross & 4 & "Cross $<$street$>$ at $<$place$>$\\
upper,take & 5 & "Take the stairs"\\
roundabout & 6 & "At the roundabout, take exit onto $<$place$>$"\\ [1ex] 
 \hline
\end{tabular}
\caption{Values assigned to each word}
\label{table:weightroute}
\end{table}

The route complexity score $S$ is then defined as :

\begin{equation}
 \mathrm{S}=\sum_{i=1..13} W_{i} / d_{i}
\label{eq:eq3}
\end{equation}

Where $W_{i}$ is the value for $word_i$ from table \ref{table:weightroute} and $d_{i}$ is the distance associated with the direction step where we found the $W_{i}$. For example,

Eq. \ref{eq:eq3}, in conjunction with the LTE flows, explains the different local dynamics observed in areas with the same FF profiles and with the same walking distance between them. In conclusion, street configuration and core activities (and not distance) are the drivers behind the FF footprints observed in these nearby locations.

\bibliographystyle{unsrt}  
\bibliography{references}  
\section*{Supplementary material} 

\subsection*{Data construction and bias analysis} \label{SupM:S1}

The probe request detected from a device does not have a one-to-one
correspondence with an individual so the initial MAC address detected at each
location must go through a cleaning and validation procedure as detailed below.
\begin{enumerate}
\item \textbf{Input}: Hashed probe requests summarised for every five-minute interval.
\item \textbf{Count Probe requests}: We separately count the number of probe requests with *randomised* and non-randomised mac addresses.
\item \textbf{Count MAC addresses}: We then remove all the MAC addresses repeating within each five-minute interval and count the unique MAC addresses separately for random and non-randomised probe requests. For example, 15 probe requests sent from the same MAC address within the same five-minute interval would be 1.
\item \textbf{Remove long dwellers}: We then remove MAC addresses detected during consecutive intervals within a half-hour period. This removes the long dwellers from being counted repeatedly over five-minute intervals to give us the filtered counts. For example, store employees' printers or mobile devices are included only once, even if they were present over multiple consecutive intervals. This is done separately for randomised and non-randomised probe requests.
\item \textbf{Adjusting local count}: We then look at the ratio between the filtered count and the corresponding total number within the non-randomised probe requests and adjust the randomised counts for each five-minute interval. We then add the filtered, non-randomised, and adjusted randomised to arrive at the final estimated counts.

\item \textbf{Impute missing values}: Finally, we remove all gaps in the data which are less than 30 minutes long. We employ Kalman Smoothing on a structural time series model to impute the missing data from the existing ones.  The implementation of this imputation methodology is detailed at \\
https://cran.r-project.org/web/packages/imputeTS/index.html

\item \textbf{Output}: The output is cleaned footfall estimates for each five-minute interval at each location.
\end{enumerate}

\subsection*{Uncertainties in Data}
 The standard deviation was 129 and 43, respectively.
These data are used as a proxy for estimating footfall at retail locations. The sensors capture signals sent by Wi-Fi-enabled devices that are present in their range. The potentially identifiable information collected on the mobile devices is hashed at the sensor level, and the data is sent to the central server via an encrypted channel for storage.
The quality of the data is affected by a series of technical limitations relating to the Wi-Fi acquisition process:
\begin{enumerate}
\item \textbf{The range of the sensor:} Since the strength of the signal from a mobile device to the Wi-Fi access point varies depending on various factors, the sensors do not have a standard signal range. In other words, the exact delineation of the signal range is different for each sensor at different times.
\item \textbf{Probing frequency:} The rate at which a mobile device probes for available Wi-Fi access points varies widely based on the manufacturer, operating system, state of the mobile device and the number of access points already known to the device.
\item \textbf{MAC address collisions:} There are a few instances (<0.01\%) of the same MAC address being reported by different mobile devices. This might be due to aggressive MAC randomisation by mobile devices and the hashing procedure being carried out twice to anonymise the data sufficiently.
\item \textbf{On-site conditions: }These devices are installed at retail points and may occasionally be disconnected from the main power, resulting in missing data for certain intervals.
\item \textbf{Post-processing:} The process of transforming probe requests into actual footfall requires a series of assumptions that can potentially lead to over or undercounting people in the results. The methodology followed for the data delivered is explained in the next section.
\end{enumerate}

\end{document}